**An Approach to Use Depletion Charges for Modifying Band Profiles for Field-Effect Transistors**

P. Xu and H. Luo*

*Department of Physics, University at Buffalo,*

*The State University of New York, NY 14260, USA*

Abstract

We present the study of using depletion charges for tailoring lateral band profiles and applying it to the promising gate-all-around field-effect transistors (GAAFET). Specifically, we introduce heavily p-type doped Si next to the channel, but outside the channel, of a transistor. They are connected to the heavily n-type doped source and drain for generating the depletion charges. The finite difference method was used for simulations and the results show significant modifications of the conduction band along the channel. The depletion charges act as built-in electrodes capable of significantly modifying the band profiles of field-effect transistors. Quantum confinement within the channel has been attempted with different approaches, such as additional electrodes and point contacts. The results presented show two aspects of this approach, namely, realizing quantum confinement in an all-Si structure and tailoring band profiles within channels to modify their transport properties.

*Corresponding Author. E-mail address: luo@buffalo.edu

The development of the semiconductor technology largely relied on reducing the dimensions of metal-oxide-semiconductor field-effect transistors (MOSFET) for decades. The growing demands on device performance led to the need to change structural configurations. These include the successful fin field-effect transistors (FinFET),[1] and gate-all-around field-effect transistors (GAAFET),[2] with increasing complexity, smaller dimensions, and issues unique to reduced dimensions addressed. With the application of artificial intelligence entering a new era, more computing power and less energy consumption are becoming more critical. For Si-based FinFET and GAAFET, overcoming short channel effects has been a major step forward,[3] but their energy consumption continues to be a long-standing challenge, particularly with the increasing density of transistors.[4] Modifying the band profile for Si-based devices has been difficult with limited compatible heterostructures that can be easily formed. There have been many previous studies investigating the use of quantized states within a channel for electron transport, a different direction for field effect transistors. This is especially important because the semiconductor technologies have already reached the dimension when quantum confinement can be significant. Some of those studies focused on Si-based structures, including the use of quantum wires,[5] additional electrodes, and quantum dots with point contacts.[6] In field-effect transistors, quantum wells were created using additional electrodes due to the lack of readily available materials or structures capable of tailoring the band profiles effectively. While quantum confinement along the channel width and the gate-field directions are easily realized, the same is much more challenging along the channel length direction to be compatible with existing technologies. In this study, we present a flexible approach capable of significantly modifying the lateral band profiles within the channels of Si-based FETs, with potential applicability to other systems.

To improve the effectiveness of the applied gate field, GAAFET structures feature a metal gate that wraps around the channel and have become one of the key focused areas in the semiconductor industry. For this study, our approach is to introduce heavily doped p-type regions surrounding the underlaps, i.e. the space between the gated region and the source/drain, sometimes referred to as spacers in other cases. The cross-section is shown in Fig. 1. To prevent leakage current, dielectric materials are typically used to entirely fill the space occupied by the p-type regions.[7] In our design, we replace part of the dielectric material with heavily doped p-type silicon, either crystalline or amorphous, in contact with the heavily n-doped source and drain, which will simply be referred to as p-regions. They are separated from the channel by thin oxide layers from the channel to prevent leakage current. The depletion charges within the p-regions can significantly alter the band profiles within the channel, such as creating barriers in and around the underlap region to achieve quantum confinement among other things. The structure shown in Fig. 1 is for the convenience of simulations, and the key factors for the proposed effects will be discussed later.

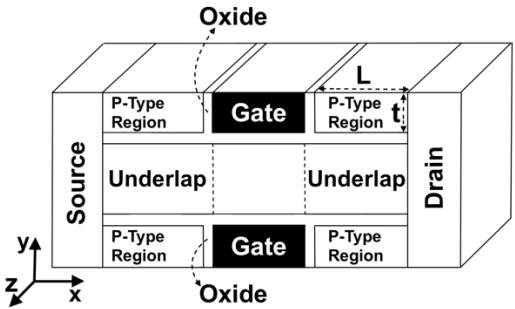

Fig. 1. A schematic diagram of a double-gate structure as the model configuration used in this work. It is an n-channel with n-doped source and drain. The focus here is primarily on the results related to the addition of the heavily doped p-regions.

The profile of the Si conduction band relative to the Fermi level is determined by several factors, such as the types of doping and doping levels in all the regions shown in Fig. 1, and the applied gate bias. For the p-regions, their band profile and surrounding areas will also depend on the dimensions of these regions. If the dimensions of the p-regions are greater than the full depletion layer thickness from the source and the drain, the highest conduction band edge in the p-regions will be close to the full energy gap of silicon above that of the n-type source/drain, similar to the situation of a simple silicon p-n junction diode. When the length of the p-regions, L, shown in Fig. 1, is reduced to less than the depletion thickness, the p-regions will be fully depleted, and their conduction bands will be lowered, due to insufficient depletion charges to raise the band edge. The goal of having these p-regions is to tailor the band profile in the underlap regions, without introducing leakage currents. The depletion charges in the p-regions, i.e., negative ions, can provide electric fields capable of elevating the conduction bands in the sandwiched underlap regions, forming barriers along the channel, as will be seen in the simulation results below.

For this study, the whole channel itself, including both the gated region and the underlaps, is lightly p-doped Si, while the source/drain are highly n-doped, as in most GAAFETs. We solve for the charge distribution, the corresponding band profile, and quantization in the channel self-consistently.  The parameters are chosen to resemble those used in the state-of-the-art GAAFET designs.  The oxide layer has a thickness of 1 nm, and the oxide material used is hafnium oxide, chosen for its high dielectric constant of 22.[8]  The channel width, in the z-direction in Fig. 1, is around 40 nm,[9] and the quantization effect is neglected in this direction.  We also neglect the effect of the gates at the ends in the width direction because of the long width, reducing a four-gate all-around structure to a two-gate structure as shown in Fig. 1, for simplicity.  This simplifies the simulations, from a 3D system to a 2D system.  To make sure that this approximation is valid, we performed a 2D simulation for the charge distribution in the y-z plane (see Fig. 1), which shows negligible effects related to the two gates at the ends in the width direction.  We are left with the charge distribution and potential energy profile in the x-y plane of the channel in Fig. 1.  The temperature is 300 K in our simulations.

For the simulation, we adopt the commonly used finite difference method to solve the coupled Poisson and Schrodinger equations, ideal for obtaining charge distributions in complex structures and quantization in the corresponding quantum wells in a self-consistent manner.[10] In the simulation, we only treat the electrons in the channel quantum mechanically because of the fact that holes do not change directly the transport properties of the device, and quantization of holes affect the overall charge distribution minimally.  Also we only treat the case at zero source/drain voltage, and we use the time-independent Schrodinger equation and ignore the coupling between the states in the channel with the continuums outside the channel.

A 2D mesh grid is used for the simulation with M points along the x-direction (channel length) and N points along the y-direction (gate field direction), with a uniform grid spacing $a$. In the finite difference method, the Hamiltonian is discretized at each mesh point $(x_i, y_j)$ in the channel in the form:

$$H\Psi_\mu(x_i, y_j) = [U(x_i, y_j) + 2t_x + 2t_y]\Psi_\mu(x_i, y_j) - t_x\Psi_\mu(x_{i-1}, y_j) - t_x\Psi_\mu(x_{i+1}, y_j) - t_y\Psi_\mu(x_i, y_{j-1}) - t_y\Psi_\mu(x_i, y_{j+1}), \quad (1)$$

where $\Psi_\mu(x_i, y_j)$ is the wavefunctions of the quantized states in the x-y plane. $U(x_i, y_j)$ is the potential energy at mesh point $(x_i, y_j)$, and $t_x = \frac{\hbar^2}{2m_x a^2}$ and $t_y = \frac{\hbar^2}{2m_y a^2}$. $m_x$ and $m_y$ are the effective masses of the electrons in the x and y directions, respectively. The effective masses are calculated from the six-fold degenerate valleys in the conduction bands of Si, with the assumption that the channel length is orientated along the [110] direction and the gate field is applied in the [001] direction, which corresponds to the orientation of the silicon substrate—a common configuration.

For the Poisson's equation, we again discretize it using the finite difference method,[11] in the form of:

$$-\frac{1}{qa^2}[U(x_{i-1}, y_j) + U(x_{i+1}, y_j) + U(x_i, y_{j-1}) + U(x_i, y_{j+1}) - 4U(x_i, y_j)]$$

$$= -\frac{\rho(x_i, y_j)}{\varepsilon_{Si}}, \quad (2)$$

where $q$ is the magnitude of the electron charge, and $\varepsilon_{Si}$ denotes the permittivity of Silicon. The term $\rho(x_i, y_j)$ represents the charge density at a specific mesh point, given by: $\rho(x_i, y_j) =$

$q[-n(x_i, y_j) + p(x_i, y_j) - N_A(x_i, y_j) + N_D(x_i, y_j)]$, where $n$ is the electron density per unit volume, $p$ the hole density, $N_A$ the acceptor ion density, and $N_D$ the donor ion density with their Fermi distributions included. These equations replace the derivatives with a system of equations, represented as a matrix of dimensions $MN \times MN$, which can be solved for the eigenstates, the charge distribution and the band profile.[10] Equations (1) and (2) are coupled because the quantization affects the charge density. Therefore, they are solved self-consistently. We first apply the Gauss-Seidel method to obtain two trial potentials,[11] which are subsequently used in a secant method iteration to obtain the final result.[12]

Another approximation made in our simulations is to neglect the voltage drop over the extremely thin oxide layer, i.e., 1 nm, and Dirichlet boundary conditions are applied around the gates.[13] Given that the hafnium oxide has a high dielectric constant of 22, the voltage drops across the thin oxide layers are expected to be small in comparison with the overall conduction band profile, shown in Fig. 2, and the effect on the middle-line band profile of the channel (shown later in Fig. 4) is expected to be negligible.

The parameters used for the result shown in Fig. 2 are the following. The heavily doped p-regions have doping concentrations of $8 \times 10^{19}\ cm^{-3}$, on the same order of magnitude compared to those typically used for highly doped source/drain. The length of the p-regions is $L = 9\ nm$, and the thickness $t = 4\ nm$, as shown in Fig. 1. The gates and the p-regions share the same thickness of $4\ nm$ purely for the convenience of simulations, which allows the system to maintain an overall rectangular shape. The gates have a length of $5\ nm$ in the x-direction (channel length direction), and the gate separation in the y-direction is also 5 $nm$. The p-type channel, including the underlap, is lightly p-type doped with a concentration of $1 \times 10^{17}\ cm^{-3}$.[14] The n-type source/drain are doped with a concentration of $1 \times 10^{20}\ cm^{-3}$. The work function of the gate material is chosen to be 4.2 eV.[15] The applied gate voltage is 0.5 V.

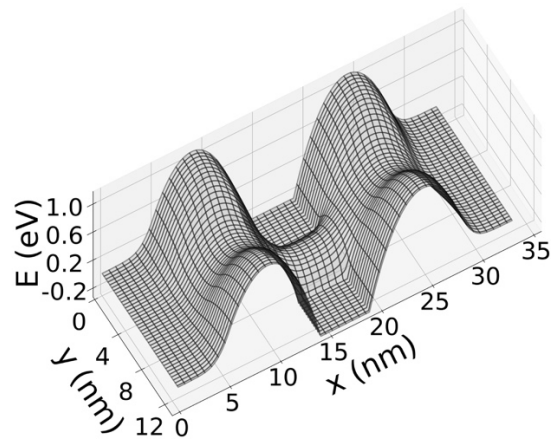

Fig. 2. A plot of the conduction band profile with a gate voltage of 0.5V, for the entire structure shown in Fig. 1, except the barriers of the oxide layers, for clarity of view.

Figure 2 shows the conduction band profile in the xy-plane, for the configuration shown in Fig. 1. The four peaks correspond to the p-regions. The flat plains between the peaks represent the energy positions of the gates. On the left and right sides are the source and drain regions. The channel is situated in the middle, running along the x-direction and sandwiched between the peaks.

The result for the middle-line along the channel, shown in Fig. 3, provides more insight about the channel and directly demonstrates the extent of the modification of the band profile and the confinement effect within the channel. Two energy barriers along the channel length direction (x-direction) are clearly seen in Fig. 3. The barrier shapes and heights can be changed as a function of both the doping concentration and the length L of the p-regions. In Fig. 3(a), the barriers in the underlaps are approximately 550 meV above the Fermi level with the parameters given earlier, close to half of the band gap of Si. In Fig. 3(b), the length of the p-regions is reduced to $8\ nm$, and the doping concentration is increased to $1 \times 10^{20}\ cm^{-3}$ to maintain a similar amount of total depletion charge, while all other parameters remain the same. The barrier height in the underlaps within the channel decreases slightly to approximately 530 meV above the Fermi level, with narrower barrier width.

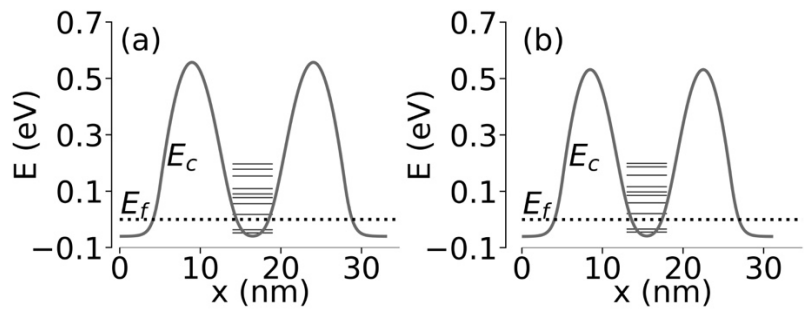

Fig. 3. Conduction band profile along the middle-line of the channel with low-lying quantized states, with a gate voltage of 0.5 V. The short solid lines are the quantized states up to 200 meV above the Fermi level, and the long-dotted line is the Fermi level. The two key parameters for the p-regions are (a) $9\ nm$ long, with a doping concentration of $8 \times 10^{19}\ cm^{-3}$, and (b) $8\ nm$ long, with a doping concentration of $1 \times 10^{20}\ cm^{-3}$, while all other parameters remain unchanged.

The results in Fig. 3 indicate that depletion charges can significantly alter the energy profile within the channel for modifying the functionality of field-effect transistors, whether to achieve quantization or to improve on-off ratios of a transistor, with more details given below.  Both the barrier heights and widths, as well as the spacings between the quantized states, can be adjusted with the parameters listed before.  The results shown are only intended to demonstrate the range of modification of the lateral band profile.  More comprehensive studies are needed to investigate transport properties with a source/drain voltage and optimization of the parameters.

Simulations show that varying the numbers of p-regions, in configurations different from the symmetrical arrangement shown in Fig. 1, does not affect the basic function of the depletion charges in individual p-regions.  One example simulated is to remove the bottom two p-regions in the structure shown in Fig. 1, which simplifies the fabrication process, and will be presented elsewhere.  Another example is to only have p-regions on either the left or the right side of the gated region shown in Fig. 1 for desired transport properties.  The basic effect remains as long as the p-regions are located in proximity to the underlaps, have sufficient depletion charges, and are well insulated from the channel.  These are similar to the conditions for electrodes. The effects of the depletion charges are in fact the same as the role of charges in floating-gates for memory devices.  Such charges also modify the band profile, producing barriers along the channel to form the on and off states of floating gate transistors.  The difference is that the depletion charges in p-regions are built-in, with minimal charging/discharging during the operation that always negatively impacts the speed.

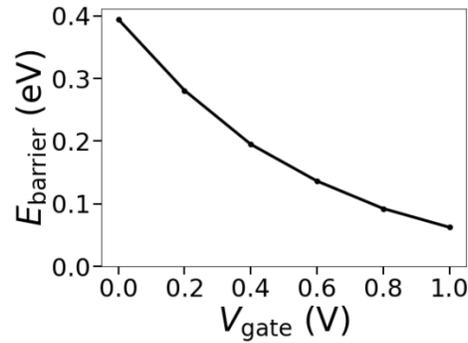

Fig. 4. The simulated result with a p-region length of 6.5 nm and a doping concentration of $8 \times 10^{19}\ cm^{-3}$, as a function of the gate bias, $V_{gate}$. It indicates that at the off state, $V_{gate}$ = 0, the channel has much higher barriers along the middle-line, which can be reduced to 60 meV once $V_{gate}$ becomes 1.0 V.

Shown in Fig. 4 is the case of p-region with length L = 6.5 nm, and a doping concentration of $8 \times 10^{19}\ cm^{-3}$. We present the barrier height, $E_{barrier}$ in the figure, along the middle-line of the channel, as a function of gate bias, $V_{gate}$. The sensitivity of the barrier height on $V_{gate}$ depends on the length of the p-region. This can be seen by comparing the results in Fig. 3 and Fig. 4 at $V_{gate}$ = 0.5 V. The barriers in Fig. 3 are high at this gate bias, whereas they are much reduced in Fig. 4 at the same bias, typical for reduced p-region length in our simulations. In other words, it is similar to the case of a weaker electrode. The results in Fig. 4 benefits on-off ratio of a field effect transistor, because at $V_{gate}$ = 0, the off state, the barrier is high, and as $V_{gate}$ moves to the on state, the barrier height is reduced significantly. This adds yet another degree of freedom for device designs.

The approach can be viewed as implanting built-in electrodes in the form of depletion charges, which can be tailored depending on the need.   It can be used for already widely used structures, such as FinFET structures to reduce the leakage current.   The width and height of the barriers within the channel shown here are only to demonstrate the range of changes in the band profile using this approach, not optimized for a particular device specification.  Another flexibility of this approach is that it can be applied to other structures and materials, such as GaAs-based structures and Si-based p-channel FETs.  When the depletion charges are located next to 2D materials, either in the substrate or an overlayer, lateral band structures can be tailored.  For widely-studied Si-based structures, depletion charges can be situated in either crystalline or amorphous Si, because both can be doped up to $1 \times 10^{20}$ cm$^{-3}$.  The ability to easily modify lateral band profiles and the reduced reliance on heterostructures make this approach complementary to the common method for band engineering in semiconductors, namely, heterostructures, which are ideal along the growth (or vertical) direction.

**References**

[1] K. Karimi, A. Fardoost and M. Javanmard, "Comprehensive Review of FinFET Technology: History, Structure, Challenges, Innovations, and Emerging Sensing Applications," Micromachines **15** (10), 1187 (2024).
[2] S. Mukesh and J. Zhang, "A Review of the Gate-All-Around Nanosheet FET Process Opportunities," Electronics **11** (21), 3589 (2022).
[3] A. Kumar and S. S. Singh, "Optimizing FinFET parameters for minimizing short channel effects," in *2016 International Conference on Communication and Signal Processing (ICCSP)*, Melmaruvathur, India, 2016, (IEEE), p. 1448-1451.
[4] V. A. Chhabria and S. S. Sapatnekar, "Impact of Self-heating on Performance and Reliability in FinFET and GAAFET Designs," in *20th International Symposium on Quality Electronic Design (ISQED)*, Santa Clara, CA, USA, 2019, (IEEE), p. 235-240.
[5] J. P. Colinge, X. Baie, V. Bayot and E. Grivei, "A silicon-on-insulator quantum wire," Solid-State Electron. **39** (1), 49-51 (1996).
[6] H. Ishikuro and T. Hiramoto, "Quantum mechanical effects in the silicon quantum dot in a single-electron transistor," Appl. Phys. Lett. **71** (25), 3691-3693 (1997).
[7] F. B. Z. Mo, C. E. Spano, Y. Ardesi, M. R. Roch, G. Piccinini and M. Vacca, "NS-GAAFET Compact Modeling: Technological Challenges in Sub-3-nm Circuit Performance," Electronics **12** (6), 1487 (2023).
[8] J. H. Choi, Y. Mao and J. P. Chang, "Development of hafnium based high-k materials—A review," Mater. Sci. Eng. R Rep. **72** (6), 97-136 (2011).
[9] S. Davies. "IBM Announces 2nm GAA-FET Technology – the Sum of "Aha!" Moments." 2021. Accessed on 2024. https://www.semiconductor-digest.com/ibm-announces-2nm-gaa-fet-technology-the-sum-of-aha-moments/.
[10] A. Sundar and N. Sarkar, "Effect of size quantization and quantum capacitance on the threshold voltage of a 2D nanoscale dual gate MOSFET," Eng. Res. Express **2** (3) (2020).
[11] R. J. LeVeque, *Finite Difference Methods for Ordinary and Partial Differential Equations: Steady-State and Time-Dependent Problems*, 1 ed. (Society for Industrial and Applied Mathematics, 2007), p. 69.
[12] M. Newman, *Computational Physics*, 1 ed. (Createspace, North Charleston, SC, 2013), p. 273.
[13] F. R. Shapiro, "The numerical solution of Poisson's equation in a pn diode using a spreadsheet," IEEE Trans. Educ. **38** (4), 380-384 (1995).
[14] N. Totorica, W. Hu and F. Li, "Simulation of different structured gate-all-around FETs for 2 nm node," Eng. Res. Express **6**, 035326 (2024).
[15] E. Elke, H. Klaus and T. Dina, "Work Function Setting in High-k Metal Gate Devices," in *Complementary Metal Oxide Semiconductor*, edited by Y. Kim Ho and N. Humaira (IntechOpen, Rijeka, 2018).